\documentclass[letterpaper, 10 pt, conference]{ieeeconf}  

\IEEEoverridecommandlockouts                              

\usepackage{amsmath} 
\usepackage{pdfpages}
\usepackage{subfigure}
\usepackage{cite}
\usepackage[bookmarks=true,colorlinks,linkcolor=black,citecolor=black,urlcolor=black]{hyperref}

\newtheorem{mythe}{\textbf{Theorem}}[section]
\graphicspath{{Graphics/}}
\title{\LARGE \bf Impedance Control of A Cable-Driven Series Elastic Actuator \\ with the 2-DOF Control Structure}

\author{Wulin Zou, Zhuo Yang, Wen Tan, Meng Wang, Jingtai Liu, Ningbo Yu* \\
\thanks{This work has been supported by the National Natural Science Foundation of China (61403215), the Natural Science Foundation of Tianjin (13JCYBJC36600) and the Fundamental Research Funds for the Central Universities.}%
\thanks{Corresponding author Assoc. Prof. Dr. Ningbo Yu is with the Institute of Robotics and Automatic Information Systems, Nankai University, and Tianjin Key Laboratory of Intelligent Robotics, Nankai University, Haihe Education Park, Tianjin 300353, China. Phone: +86 (0)22 2350 3960 ext. 801, Email: nyu@nankai.edu.cn.}%
\thanks{Mr. Wulin Zou, Mr. Zhuo Yang, Mr. Wen Tan, Mr. Meng Wang and Prof. Dr. Jingtai Liu are with the Institute of Robotics and Automatic Information Systems, Nankai University, and Tianjin Key Laboratory of Intelligent Robotics, Nankai University, Haihe Education Park, Tianjin 300353, China.}%
}

\begin{document}
	
\maketitle
\thispagestyle{empty}
\pagestyle{empty}

\begin{abstract}

Series elastic actuators (SEAs) are growingly important in physical human-robot interaction (HRI) due to their inherent safety and compliance. Cable-driven SEAs also allow flexible installation and remote torque transmission, etc. However, there are still challenges for the impedance
control of cable-driven SEAs, such as the reduced bandwidth caused by the elastic component, and the performance balance between reference tracking and robustness. In this paper, a velocity sourced cable-driven SEA has been set up. Then, a stabilizing 2 degrees of freedom (2-DOF) control approach was designed to separately pursue the goals of robustness and torque tracking. Further, the impedance control structure for human-robot interaction was designed and implemented with a torque compensator. Both simulation and practical experiments have
validated the efficacy of the 2-DOF method for the control of cable-driven SEAs.

\end{abstract}

\section{Introduction}

Physical human-robot interaction has become significantly more important in recent years and is now a major focus in robotics and control. Safety and compliance should be considered in its design process. SEAs, first introduced in\cite{Pratt1995}, show a number of benefits, such as great shock tolerance, accurate and stable force control, safety and energy storage, etc. Thus, SEAs have been widely employed in various physical human-robot interaction applications to improve safety and compliance\cite{Robinson1999,Kong2012,Yu2013}.

Cables with low weight to length ratio can change the force direction intentionally and easily, enable torque transmission to a remote distance with less energy loss and space occupation \cite{Chapuis2006,Caverly2015}, and allow detachment of the actuation motor from the robot frame \cite{Veneman2006}. Besides, cable actuation can be a safe solution in human-robot interaction due to its unidirectional force constraint and property of breaking when the tension exceeding a certain threshold. Thus, Many physical human-robot interaction applications have adopted cable actuation for force or torque transmission~\cite{Veneman2006,Oblak2011,Sergi2015}. A cable-driven SEA can combine the advantages of SEA and cable actuation.

SEAs are initially designed to output a desired torque. Different torque control methods have been developed to achieve high performance for different SEAs. In \cite{Robinson2000}, Robinson introduced a PID control method to explicitly illustrate the concept and performance of SEAs. Subsequently, a lot of researchers have made important contribution to PID based SEA control~\cite{Veneman2006,Vallery2008,Oblak2011,Sergi2015,Wyeth2008}. In~\cite{Kong2012,Yoo2015,Lu2015}, disturbance observer (DOB)-based control methods have also been adopted to enhance robustness. Wyeth proposed a cascaded torque control method with an inner velocity loop to overcome  nonlinearity problems~\cite{Wyeth2008}.

Impedance control is a fundamentally important strategy to regulate the system compliance, especially for physical human-robot interaction. The impedance value in one sense represents the dynamic relationship between the interaction force/torque and the motion. Different impedance values are needed for various interaction tasks with different people. A torque-impedance cascaded control structure has been applied \cite{Vallery2008,Sergi2015} and achieved good performance.

Horowitz revealed that the conventional 1-DOF control configuration was unable to cope with the two problems of achieving a desired tracking and attaining good robustness at the same time, but the 2-DOF control structure could achieve these two goals simultaneously~\cite{Horowitz1963}. The reason is that the 2-DOF control structure provides an independent way to optimize disturbance/noise rejection and improve tracking response. Theories and practice about the 2-DOF control can be found in~\cite{Youla1985,Qiu2010,Huang2015}. Detailed simulation results of torque control using the 2-DOF control method for a cable-driven SEA have been presented in our previous work~\cite{ZouWulin2016}.

In this paper, we design the torque controller using the 2-DOF control structure to track torque reference signal, eliminate disturbance and noise, and further, to improve bandwidth. The impedance control structure for human-robot interaction is then designed based on the 2-DOF torque
controller and implemented with a compensator.

The paper is organized as follows. Section \ref{section2} describes the hardware platform and modeling of the designed cable-driven SEA. Details of the controller design are given in Section \ref{section3}. Simulation, practical experiments and results are presented in Section \ref{section4}. Finally, Section \ref{section5} concludes the paper.

\section{The Cable-driven Series Elastic Actuator}\label{section2}

\subsection{Mechanical Structure}

\begin{figure}[h]
\centering
\includegraphics[width=0.95\columnwidth]{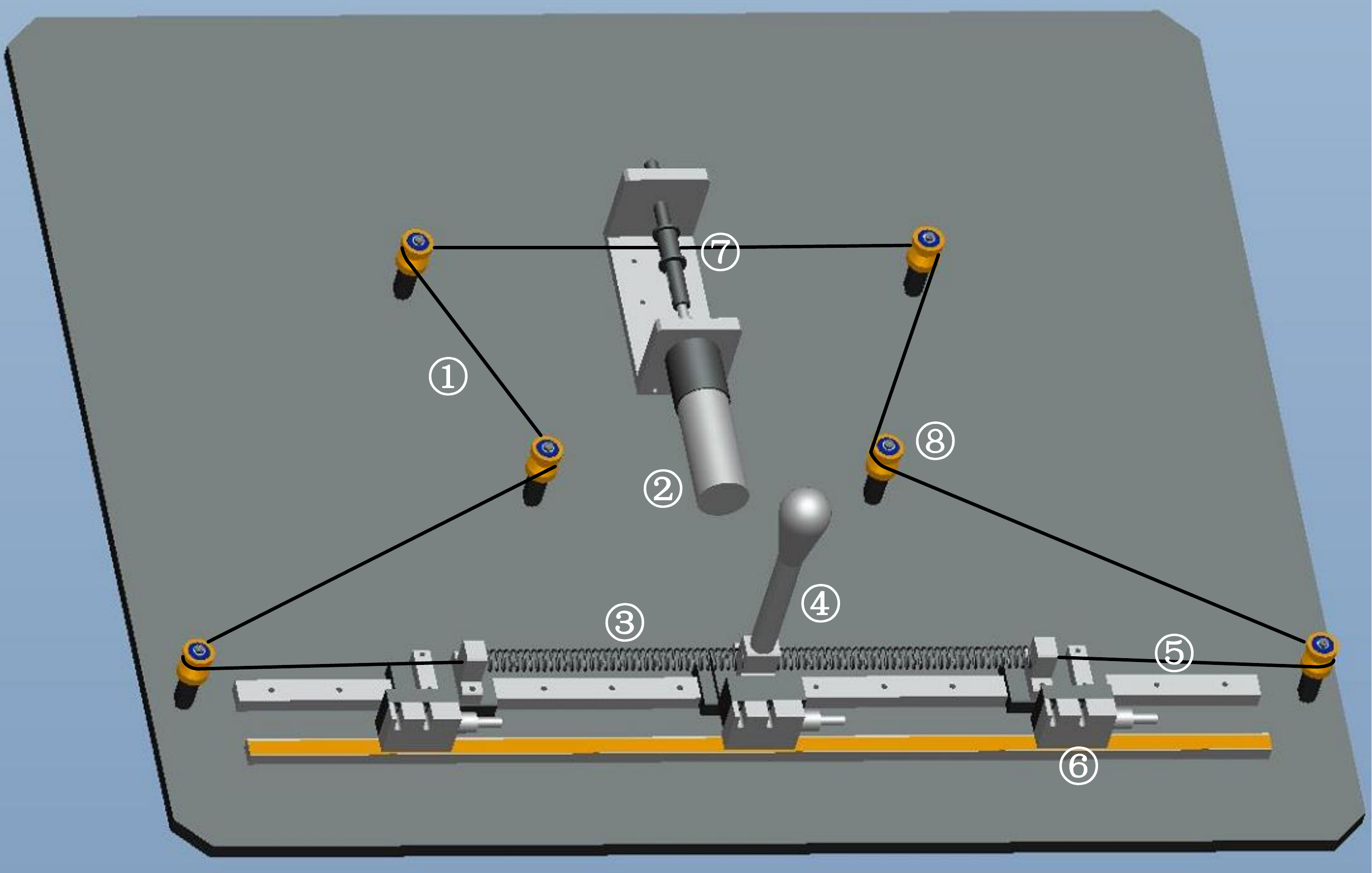}
\caption{The structure of the cable-driven SEA platform: (1) cable for transmission; (2) DC motor, including a reduction gear head and a rotary encoder; (3) linear springs as the elastic components; (4) handle for human-robot interaction; (5) sliding guide; (6) magnetic linear encoder to measure the displacement of the slider, including a reading head and a magnetic tape attached parallely to the sliding guide; (7) cable winch; (8) pulleys for cable transmission and redirection.}\label{mechanical_design}
\end{figure}

\begin{figure}[h]
\centering
\includegraphics[width=0.95\columnwidth]{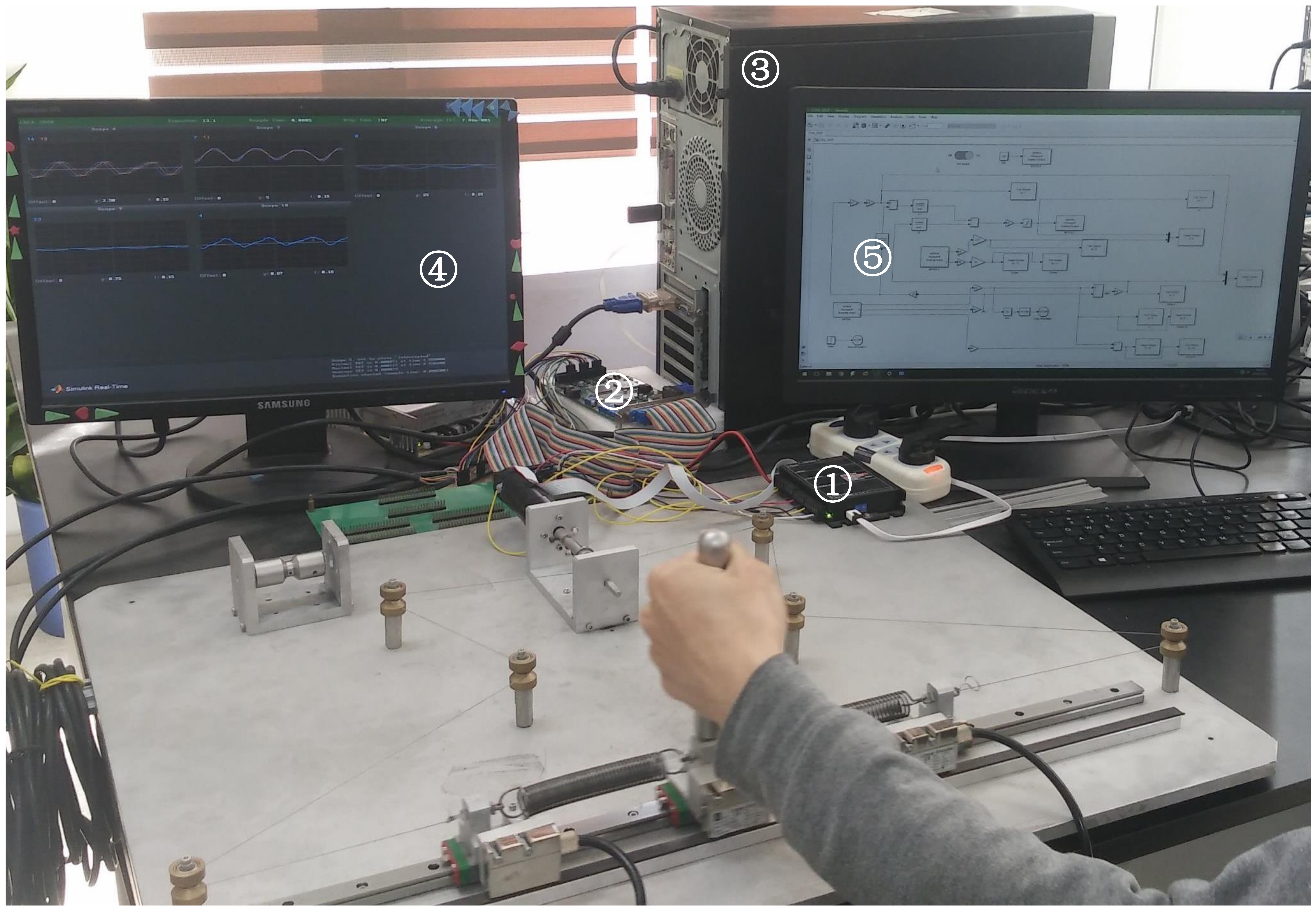}
\caption{The prototype of the cable-driven SEA platform: (1) servo controller for the DC motor; (2) data acquisition board; (3) Simulink Real-Time target computer; (4) scopes to display the data in real-time; (5) host computer.}\label{prototype}
\end{figure}

A cable-driven SEA is illustrated in Fig.~\ref{mechanical_design}. Its realized prototype is shown in Fig.~\ref{prototype}. It is a 1-DOF platform for physical human-robot interaction~\cite{Yu2014}.

A DC brush rotary motor (Maxon RE30 60W 24V) is used as the velocity source. The motor has a MR-228452 500CPT incremental encoder mounted and a GP32A 14:1 planetary gear head installed on its output shaft. The motor velocity is managed by a servo controller (Escon 50/5 409510). Both sides of the handle are connected to the motor in a cable-spring series structure. The handle is mounted at a slider and can move along the guide by cable transmission or human arm motion. The two cables are redirected by the pulleys and wrapped around the winch. The two springs have an initial length of 87.7 mm, and can be extended to the maximal length of 224.7 mm. They are pretensioned to half of their maximum displacement range. There is also a slider at each conjunction of the springs and the cables. The displacement of the handle and the deformations of the two springs are measured by three magnetic linear encoders (MLS105) mounted at each slider, with the precision of 5 $\mu$m. The length of the sliding guide is 60 cm, which is sufficient for the linear movement range of the upper limb.

Control of the actuator and data acquisition of the sensors are realized by a MATLAB/Simulink Real-Time Target System running on a standard computer. A data acquisition board (HUMUSOFT MF634 PCI-Express multifunction I/O card) is inserted into the PCI-Express x1 slot of the target computer to send the velocity command as analog signal to the motor servo controller and receive the encoder readings. The control algorithm is implemented with MATLAB/Simulink, compiled in the host computer, downloaded to the target computer and runs there in real-time.

\subsection{System Modeling}

\begin{figure}[!htbp]
\centering
\includegraphics[width=0.9\columnwidth]{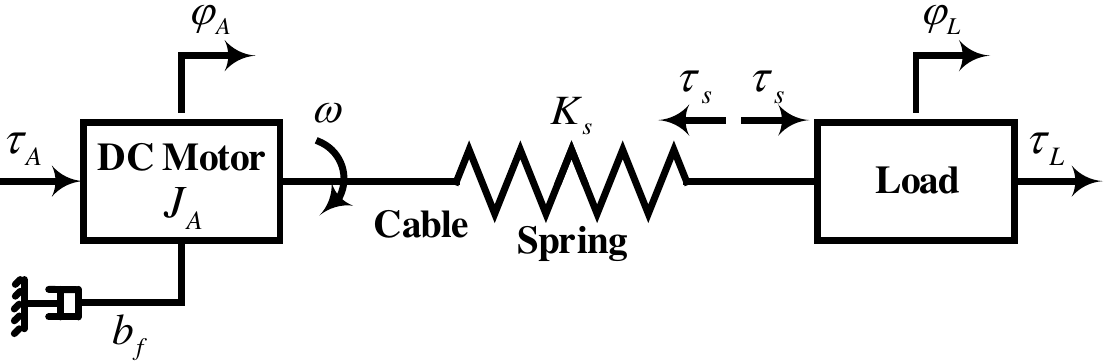}
\caption{Schematic diagram of the cable-driven SEA system}\label{SEA}
\end{figure}

The schematic diagram is shown in Fig.~\ref{SEA}. The dynamics of the interaction system can be described by the following two equations:
\begin{equation}\label{eq1}
  {\tau _A} - {\tau _L} = {J_A}{\ddot \varphi _A} + {b_f}{\dot \varphi _A}
\end{equation}
\begin{equation}\label{eq2}
  {\tau _L} = {K_s}({\varphi _A} - {\varphi _L})
\end{equation}
Where, the variable $\tau_L$, which equals to interaction torque $\tau_s$, is the output torque acting on the load. $\varphi_A$ and $\varphi_L$ are the displacements of the motor and load.  $K_s$ is the equivalent stiffness of the two springs. $J_A$ represents the reflected inertia of the motor. $b_f$ is the reflected damping of the viscous friction of the system. All parameters and variables are converted from translational to rotational motion with respect to the cable winch.

Here, we let $\varphi_L=0$, substitute (\ref{eq2}) into (\ref{eq1}), then, $\tau_L(s)$ can be obtained as:
\begin{equation*}
  {\tau _L}(s) = \frac{{{K_s}}}{{{J_A}{s^2} + {b_f}s + {K_s}}}{\tau _A}(s)
\end{equation*}

\begin{figure}[!htbp]
\centering
\includegraphics[width=0.9\columnwidth]{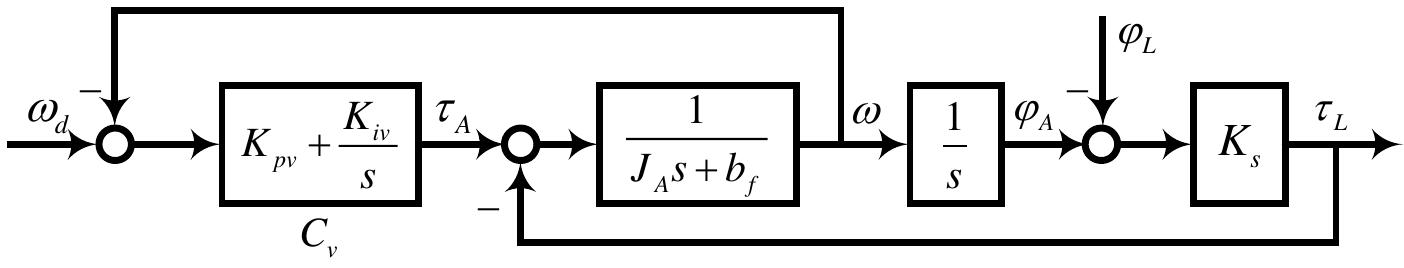}
\caption{Schematic diagram of a velocity sourced cable-driven SEA}\label{velocity_sourced}
\end{figure}

To design a well performing SEA, the DC motor is used as the velocity source of the actuator. The reason is that velocity control can overcome some undesirable effects caused by motor internal disturbance. The schematic diagram of a velocity sourced cable-driven SEA is shown in Fig.~\ref{velocity_sourced}. The velocity feedback regulated by a well tuned PI controller forms a stable closed loop so that the DC Motor system can track the reference velocity signal quickly and accurately. 

Then, the dynamic relationship between $\omega_d(s)$ and $\tau_L(s)$ can be written as:
\begin{equation*}
{\tau _L}(s) = P(s){\omega _d}(s) + G(s){\varphi _L}(s),
\end{equation*}
where:
\begin{equation}\label{eq3}
P(s) = \frac{{{K_s}{K_{pv}}s + {K_s}{K_{iv}}}}{{{J_A}{s^3} + ({b_f} + {K_{pv}}){s^2} + ({K_s} + {K_{iv}})s}}
\end{equation}
and
\begin{equation*}
G(s) = -\frac{{{J_A}{K_s}{s^2} + ({K_s}{b_f} + {K_s}{K_{pv}})s + {K_s}{K_{iv}}}}{{{J_A}{s^2} + ({b_f} + {K_{pv}})s + ({K_s} + {K_{iv}})}}.
\end{equation*}

\section{Controller Design}\label{section3}

\subsection{Parameterization of All Stabilizing 2-DOF Controllers}\label{subsection_2-DOF}

\begin{figure}[!htbp]
\centering
\includegraphics[width=0.6\columnwidth]{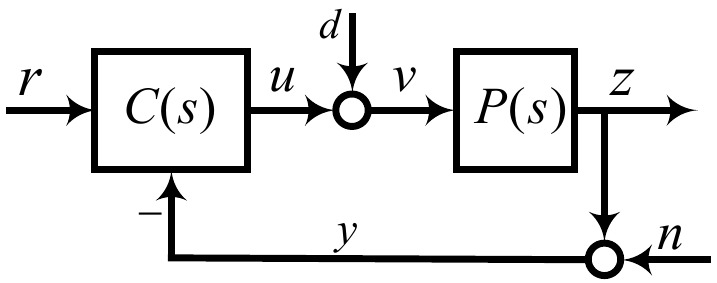}
\caption{The 2-DOF control configuration}\label{2-DOF}
\end{figure}

A closed loop system using the 2-DOF approach to stabilize the plant $P(s)$ is shown in Fig.~\ref{2-DOF}~\cite{Qiu2010}. It has two inputs $r$ and $y$ instead of one input in the 1-DOF control structure. The controller $C(s)$ can be denoted as 
$C(s) = \left[ {\begin{array}{*{20}{c}}
{{C_1}(s)}&{{C_2}(s)}
\end{array}} \right]$.
It can independently deal with the tracking and disturbance/noise rejection problem for any $P(s)$ with $P(0)\ne0$.

The four transfer functions from $r$ to the internal variables $u,v,y,z$ are:
\begin{equation}\label{eq4}
\left[ {\begin{array}{*{20}{c}}
\displaystyle {\frac{{{C_1}(s)}}{{1 + P(s){C_2}(s)}}} \vspace{2mm}\\
\displaystyle {\frac{{{C_1}(s)}}{{1 + P(s){C_2}(s)}}} \vspace{2mm}\\
\displaystyle {\frac{{P(s){C_1}(s)}}{{1 + P(s){C_2}(s)}}} \vspace{2mm}\\
\displaystyle {\frac{{P(s){C_1}(s)}}{{1 + P(s){C_2}(s)}}}
\end{array}} \right]
\end{equation}

The eight transfer functions from $d,n$ to $u,v,y,z$ are:
\begin{equation}\label{eq5}
\left[ {\begin{array}{*{20}{ll}}
\displaystyle
{\frac{{ - P(s){C_2}(s)}}{{1 + P(s){C_2}(s)}}}& \displaystyle {\frac{{ - {C_2}(s)}}{{1 + P(s){C_2}(s)}}} \vspace{2mm}\\
\displaystyle
{\frac{1}{{1 + P(s){C_2}(s)}}}&\displaystyle {\frac{{ - {C_2}(s)}}{{1 + P(s){C_2}(s)}}} \vspace{2mm}\\
\displaystyle
{\frac{{P(s)}}{{1 + P(s){C_2}(s)}}}& \displaystyle {\frac{1}{{1 + P(s){C_2}(s)}}} \vspace{2mm}\\
\displaystyle
{\frac{{P(s)}}{{1 + P(s){C_2}(s)}}}& \displaystyle {\frac{{ - P(s){C_2}(s)}}{{1 + P(s){C_2}(s)}}}
\end{array}} \right]
\end{equation}

According to (\ref{eq5}), the transfer dynamics from $d(t),n(t)$ to $u(t),v(t),y(t),z(t)$ are determined by the controller $C_2(s)$. Once the controller $C_2(s)$ has been designed, the transfer dynamics from $r(t)$ to $u(t),v(t),y(t),z(t)$  will only depend on the controller $C_1(s)$. 

Let
  \[P(s) = \frac{{b(s)}}{{a(s)}} = \frac{{{b_1}{s^{n - 1}} +  \cdots  + {b_n}}}{{{a_0}{s^n} + {a_1}{s^{n - 1}} +  \cdots  + {a_n}}},\]
where $a(s)$ and $b(s)$ are coprime and $a_0\ne0$.
Let
\[{C_0}(s) = \frac{{q(s)}}{{p(s)}} = \frac{{{q_0}{s^m} + {q_1}{s^{m - 1}} +  \cdots  + {q_m}}}{{{p_0}{s^m} + {p_1}{s^{m - 1}} +  \cdots  + {p_m}}},\]
where $p(s)$ and $q(s)$ are coprime and $p_0\ne0$, be any stabilizing controller, i.e., $c(s) = a(s)p(s) + b(s)q(s)$ is stable. Factorize $c(s)$ as $c(s)=f(s)h(s)$ such that deg $f(s)=n$ and deg $h(s)=m$. Let
\[M(s) = \frac{{a(s)}}{{f(s)}},N(s) = \frac{{b(s)}}{{f(s)}},X(s) = \frac{{p(s)}}{{h(s)}},Y(s) = \frac{{q(s)}}{{h(s)}}.\]
Then, $M(s),N(s),X(s)$ and $Y(s)$ are all stable transfer functions satisfying
\[P(s) = \frac{{N(s)}}{{M(s)}},{C_0}(s) = \frac{{Y(s)}}{{X(s)}},M(s)X(s) + N(s)Y(s) = 1.\]
\begin{mythe}
The set of all 2-DOF controllers that give stable closed-loop systems is
\begin{equation*}
\left\{ {C(s) = \left[ {\begin{array}{*{20}{c}}
\displaystyle
{\frac{{{Q_1}(s)}}{{X(s) - N(s){Q_2}(s)}}}&
\displaystyle
{\frac{{Y(s) + M(s){Q_2}(s)}}{{X(s) - N(s){Q_2}(s)}}}
\end{array}} \right]} \right\},
\end{equation*}
where $Q_1(s)$ and $Q_2(s)$ are two arbitrary stable systems~\cite{Youla1985}.
\end{mythe}

If one stabilizing controller was found, then all other stabilizing controllers could be obtained. Systematic design for stabilizing 2-DOF $H_2$-optimal controller is given as the following steps \cite{Huang2015}:
\begin{description}
  \item[Step 1]: Denote $P(s)$ as the form
  \[P(s) = \frac{{b(s)}}{{a(s)}} = \frac{{{b_1}{s^{n - 1}} +  \cdots  + {b_n}}}{{{a_0}{s^n} + {a_1}{s^{n - 1}} +  \cdots  + {a_n}}},\]
  where $a(s)$ and $b(s)$ are coprime and $a_0\ne0$.
  \item[Step 2]: Find a stable polynomial $d_\rho(s)$ called spectral factor such that
  \[{\rho ^2}a( - s)a(s) + b( - s)b(s) = {d_\rho }( - s){d_\rho }(s),\]
  where $\rho$ is a positive number used to give a relative weight to $u(t)$ and tracking error $e(t)$.
  \item[Step 3]: Find a stable polynomial $d_{\lambda ,k}(s)$ such that
  \[{k^2}a( - s)a(s) + {\lambda ^2}b( - s)b(s) = {d_{\lambda ,k}}( - s){d_{\lambda ,k}}(s),\]
  where $\lambda$ is a positive number used to give a relative weight to $d(t)$ and $r(t)$ and $k$ is a positive number used to give a relative weight to $n(t)$ and $r(t)$.
  \item[Step 4]: The feedback part of the optimal controller ${C_2}(s) = {q(s)}/{p(s)}$ with the same order as the plant is the unique strictly proper and type 0 pole-placement controller such that
      \[a(s)p(s) + b(s)q(s) = {d_\rho }(s){d_{\lambda ,k}}(s)\]
  \item[Step 5]: The feedforward part of the optimal controller is
  \[{C_1}(s) = \frac{{{d_\rho }(0)}}{{b(0)}}\frac{{{d_{\lambda ,k}}(s)}}{{p(s)}}.\]
\end{description}

If $\rho<1$, the system will achieve better tracking performance with more energy, vice versa; If $\lambda > k$, the system will achieve better disturbance rejection, otherwise, it achieves better noise rejection. It makes a tradeoff between reference tracking, disturbance rejection and energy consumption minimization by adjusting the parameters $\rho$, $\lambda$ and $k$ to receive a desired performance.

\subsection{Torque Control with the 2-DOF Structure}

\begin{figure}[!htbp]
\centering
\includegraphics[width=\columnwidth]{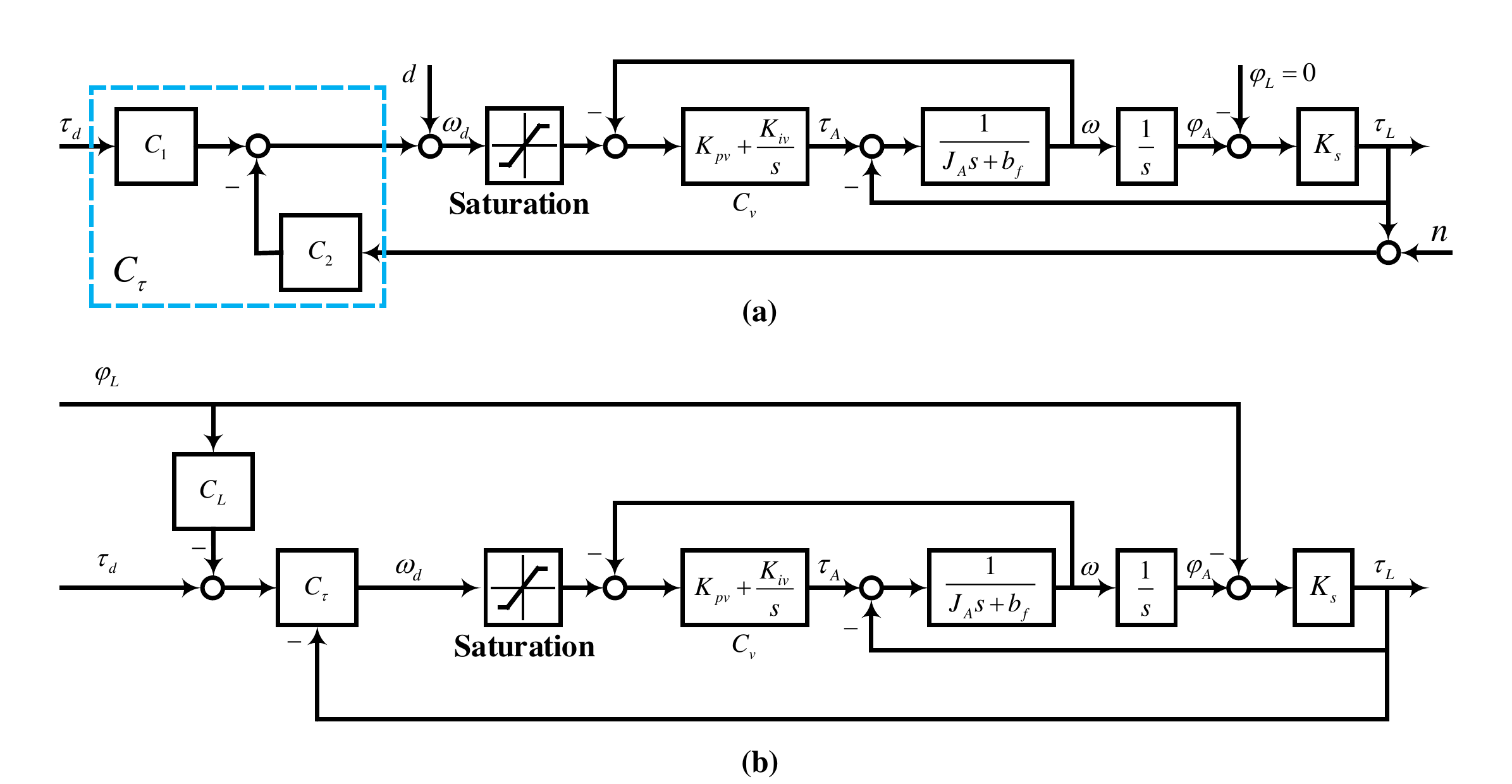}
\caption{Torque cascaded control principle based on the 2-DOF method. $C_\tau$ consisting of $C_1$ and $C_2$ denotes the 2-DOF controller. Disturbance $d$, noise $n$ and motor saturation (threshold is $\pm 50$ rad/s) are all taken into consideration. (a) The torque controller is designed under the condition of load fixed ($\varphi_L(s)=0$). (b) An additional compensator $C_L(s)$ is designed to minimize the effect caused by interaction force. }\label{torque_control}
\end{figure}

The structure and its construction process for torque control with a 2-DOF controller and a feedforward compensator are presented in Fig.~\ref{torque_control}. Firstly, a 2-DOF controller is designed to track the torque reference when the load is fixed as shown in Fig.~\ref{torque_control} (a). If the influence caused by human-robot interaction is taken into consideration, that is to say, $\varphi_L(s)$ is not equal to zero, then, $\tau_L(s)$ can be obtained as:
\begin{equation}\label{eq7}
  {\tau _L}(s) = {G_1}(s){\tau _d}(s) + {G_2}(s){\varphi _L}(s)
\end{equation}
where:
\begin{equation*}
  {G_1}(s) = \frac{{P(s){C_1}(s)}}{{1 + P(s){C_2}(s)}}, {G_2}(s) = \frac{{G(s)}}{{1 + P(s){C_2}(s)}}
\end{equation*}

Under these conditions, the output torque $\tau_L(s)$ can't track the desired torque $\tau_d(s)$ very well, because the system always has steady state error caused by $\varphi_L(s)$. However, this error can be eliminated by a feedforward compensator $C_L(s)$ presented in Fig.~\ref{torque_control}b. Let:
\begin{equation*}
  {C_L}(s) = {G_2}(s)
\end{equation*}
Then, there is:
\begin{equation}\label{eq8}
\begin{aligned}
  {\tau _L}(s) &= {G_1}(s)[{\tau _d}(s) - {C_L}(s){\varphi _L}(s)] + {G_2}(s){\varphi _L}(s) \\
  &= {G_1}(s)[{\tau _d}(s) - {G_2}(s){\varphi _L}(s)] + {G_2}(s){\varphi _L}(s) \\
  &= {G_1}(s){\tau _d}(s) + [1 - {G_1}(s)]{G_2}(s){\varphi _L}(s)
   \end{aligned}
\end{equation}
\begin{figure}[!htbp]
	\centering
	\includegraphics[width=\columnwidth]{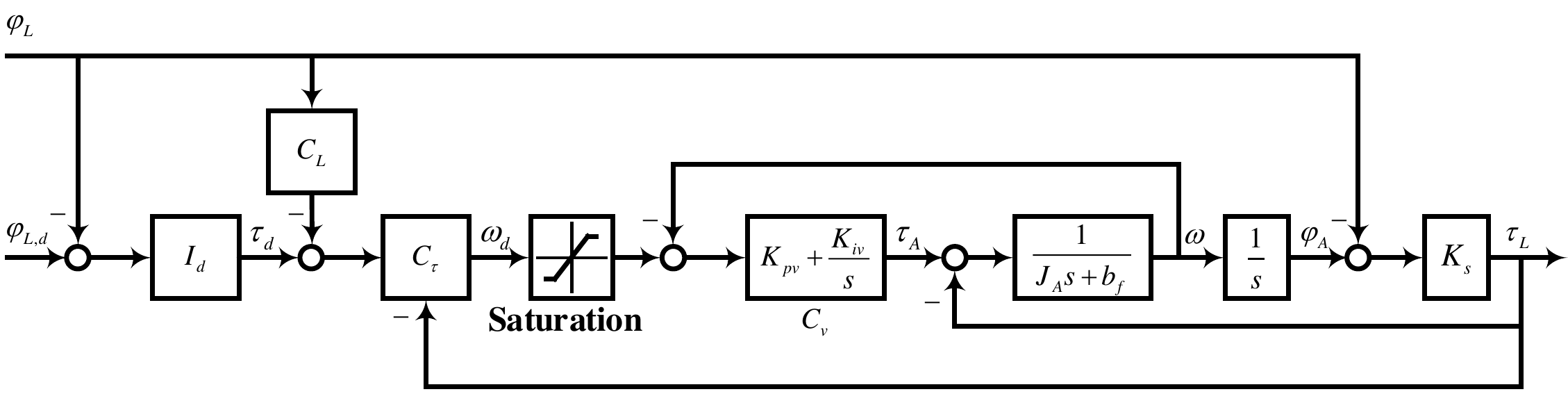}
	\caption{Impedance control for the cable-driven SEA. $I_d$ is the desired virtual impedance, and $\varphi_{L,d}$ is system reference position.}\label{impedance_control}
\end{figure}

If the torque controller $C_\tau(s)$ is ideal ($G_1(s)\simeq1$), (\ref{eq8}) is reduced as ${\tau _L}(s) \simeq {\tau _d}(s)$. Therefore, exact torque control is achieved with a 2-DOF controller $C_\tau(s)$ and a feedforward compensator $C_L(s)$. Also, the disturbance $d$ and noise $n$ are rejected by the 2-DOF controller $C_\tau(s)$.

\subsection{Impedance Control for Human-Robot Interaction}

Fig.~\ref{impedance_control} shows the block chart of the cable-driven SEA embedded in an impedance control loop. The impedance controller regulates the relation between handle motion $\varphi_L$ and resulting torque $\tau_L$. There, we only focus on virtual stiffness control, and ignore virtual inertia, virtual damping from virtual impedance. That is to say, $I_d$ is the desired virtual stiffness, which is proportional to $K_s$. For the applied handle motion, it specifies the value of desired torque $\tau_d$. The torque controller implements that output torque $\tau_L$ tracks the desired torque $\tau_d$ satisfactorily. If ${\tau _L}(s) \simeq {\tau _d}(s)$, then,
\[{\tau _L} \simeq {\tau _d} = {I_d}\left( {{\varphi _{L,d}} - {\varphi _L}} \right).\]
When the system reference point is set to zero, there is
\[{\tau _L} \simeq {\tau _d} = - {I_d}{\varphi _L}.\]

\section{Experiments and Results}\label{section4}

\subsection{Simulation and Results}

The cable-driven SEA parameters shown in Fig.~\ref{prototype} are listed in Table \ref{table1}. $J_A$ and $b_f$ are identified using MATLAB System Identification Toolbox. The velocity controller parameters $K_{pv},K_{iv}$ were tuned as 0.0457 Nm/(rad/s) and 1.3455 Nm/(rad/s).
\begin{table}[!htbp]
\caption{Cable-driven SEA parameters}
\label{table1}
\begin{center}
\renewcommand{\arraystretch}{2}
\begin{tabular}{ccc}
\hline \hline
Reflected total inertia & $J_A$ & $6.90 \times {10^{ - 4}}{\rm{kg}} \cdot {{\rm{m}}^{\rm{2}}}$\\
\hline
Coefficient of viscous friction & $b_f$ & ${\rm{0}}{\rm{.0059Nm/(rad/s)}}$\\
\hline
Stiffness of double spring & $K_s$ & $2 \times 0.0242{\rm{Nm/rad}}$\\
\hline
Radius of cable winch & $r$ & $7.25{\rm{mm}}$\\
\hline
Ratio of gear head & $K_g$ & $14:1$\\
\hline \hline
\end{tabular}
\end{center}
\end{table}

Substituting those parameters into (\ref{eq3}), there is:
\begin{equation*}
  P(s) = \frac{{{\rm{3}}{\rm{.204s + 94}}{\rm{.34}}}}{{{s^3} + 74.88{s^2} + 2021s}}
\end{equation*}

According to the design procedures in Section \ref{subsection_2-DOF}, the parameters $\rho,\lambda,k$ determining the 2-DOF controller were chosen as 0.0005, 1 and 1. Then, the 2-DOF torque controller for the cable-driven SEA is given as:
\begin{equation*}
{\small
\begin{aligned}
  &{C_1}(s) = \frac{{6.90 \cdot {{10}^{ - 4}}{s^3} + 0.0517{s^2} + 1.40s + 0.0651}}{{3.45 \cdot {{10}^{ - 7}}{s^3} + 5.07 \cdot {{10}^{ - 5}}{s^2} + 0.00346s + 0.0651}}\\
  &{C_2}(s) = \frac{{3.22 \cdot {{10}^{ - 5}}{s^2} + 0.00241s + 0.0651}}{{3.45 \cdot {{10}^{ - 7}}{s^3} + 5.07 \cdot {{10}^{ - 5}}{s^2} + 0.00346s + 0.0651}}
\end{aligned}
}
\end{equation*}

\begin{figure}[!htbp]
	\centering
	\includegraphics[width=\columnwidth]{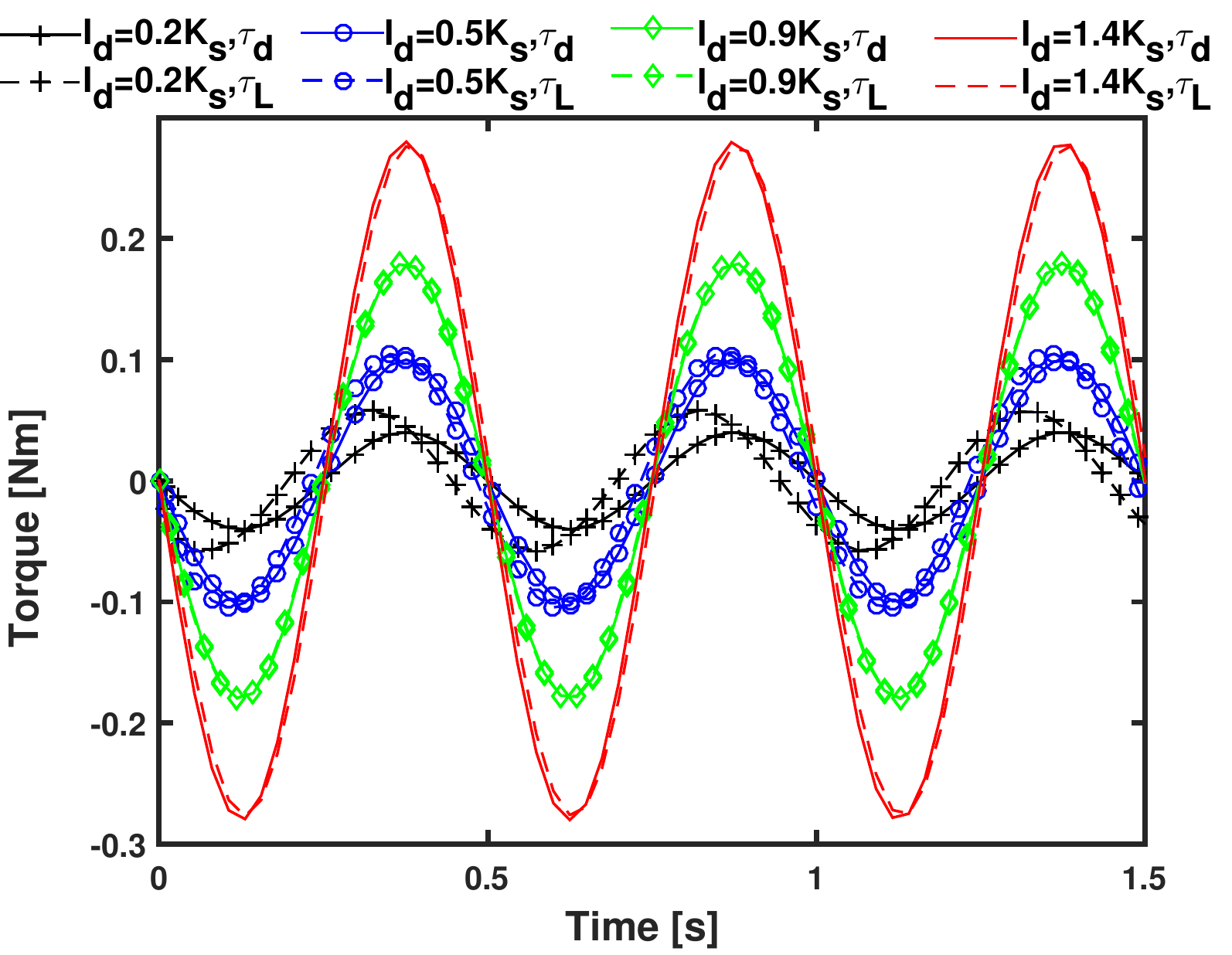}
	\caption{Simulation results of impedance control for different virtual impedances.}\label{simulation_impedance}
\end{figure}

The simulation results of impedance control presented in Fig.~\ref{impedance_control} were shown in Fig.~\ref{simulation_impedance}. System reference point $\varphi_{L,d}$ was set to zero and the handle motion caused by upper limb was set as a sinusoidal signal with a frequency of 2Hz. From the results, the output torque $\tau_L$ tracked the desired torque $\tau_d$ satisfactorily in different virtual stiffnesses changing from $0.2K_s$ to $1.4K_s$.

\subsection{Experiments and Results}

The realized prototype of the cable-driven SEA in Fig.~\ref{prototype} was used to further verify the performance of the 2-DOF torque  controller and the impedance control.

\begin{figure}[!htbp]\centering
	\begin{minipage}{\columnwidth}
		\centering
		\subfigure[]{\includegraphics[width=\columnwidth]{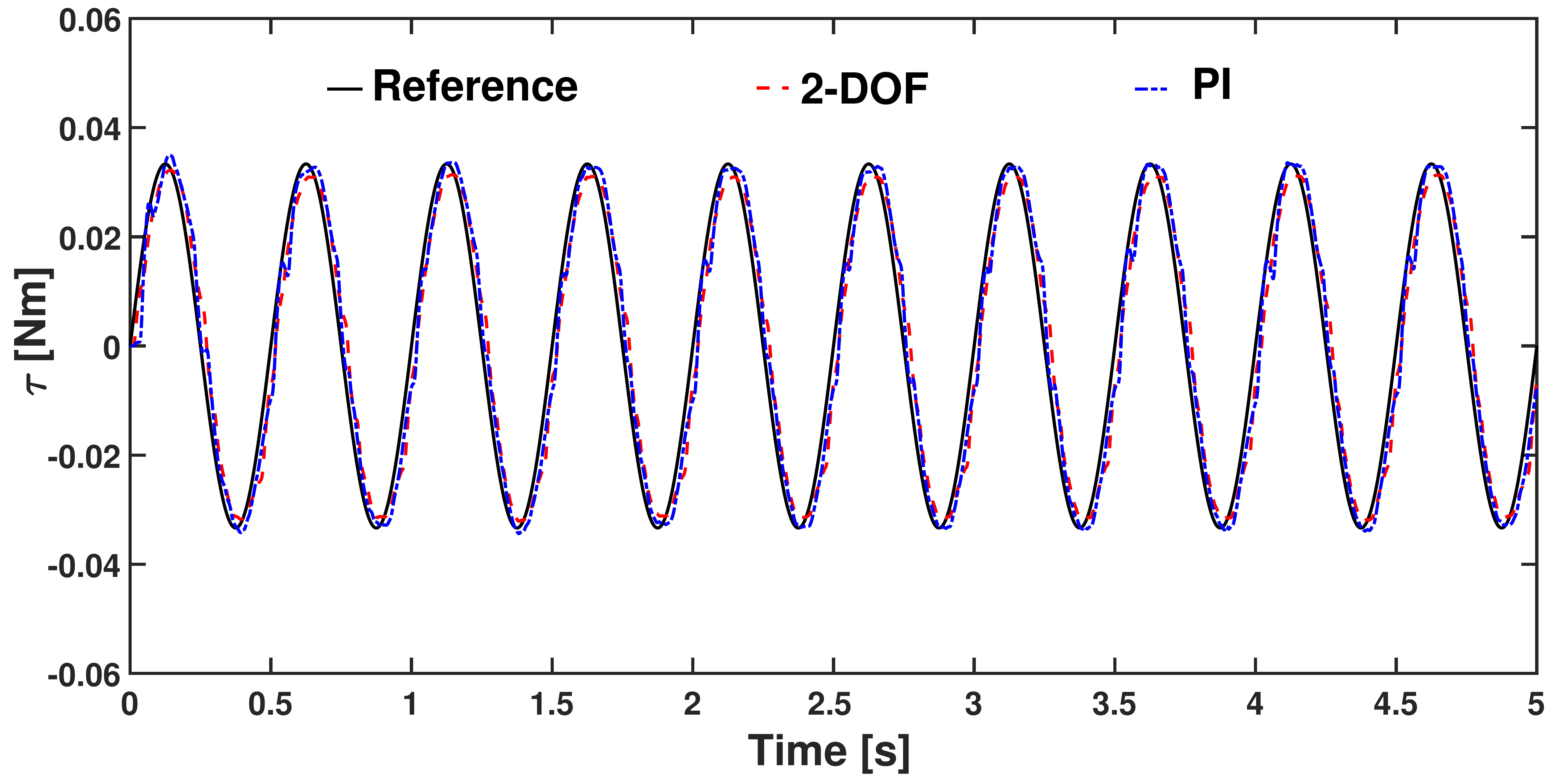}}
	\end{minipage}
	\begin{minipage}{\columnwidth}
		\centering
		\subfigure[]{\includegraphics[width=0.47\columnwidth]{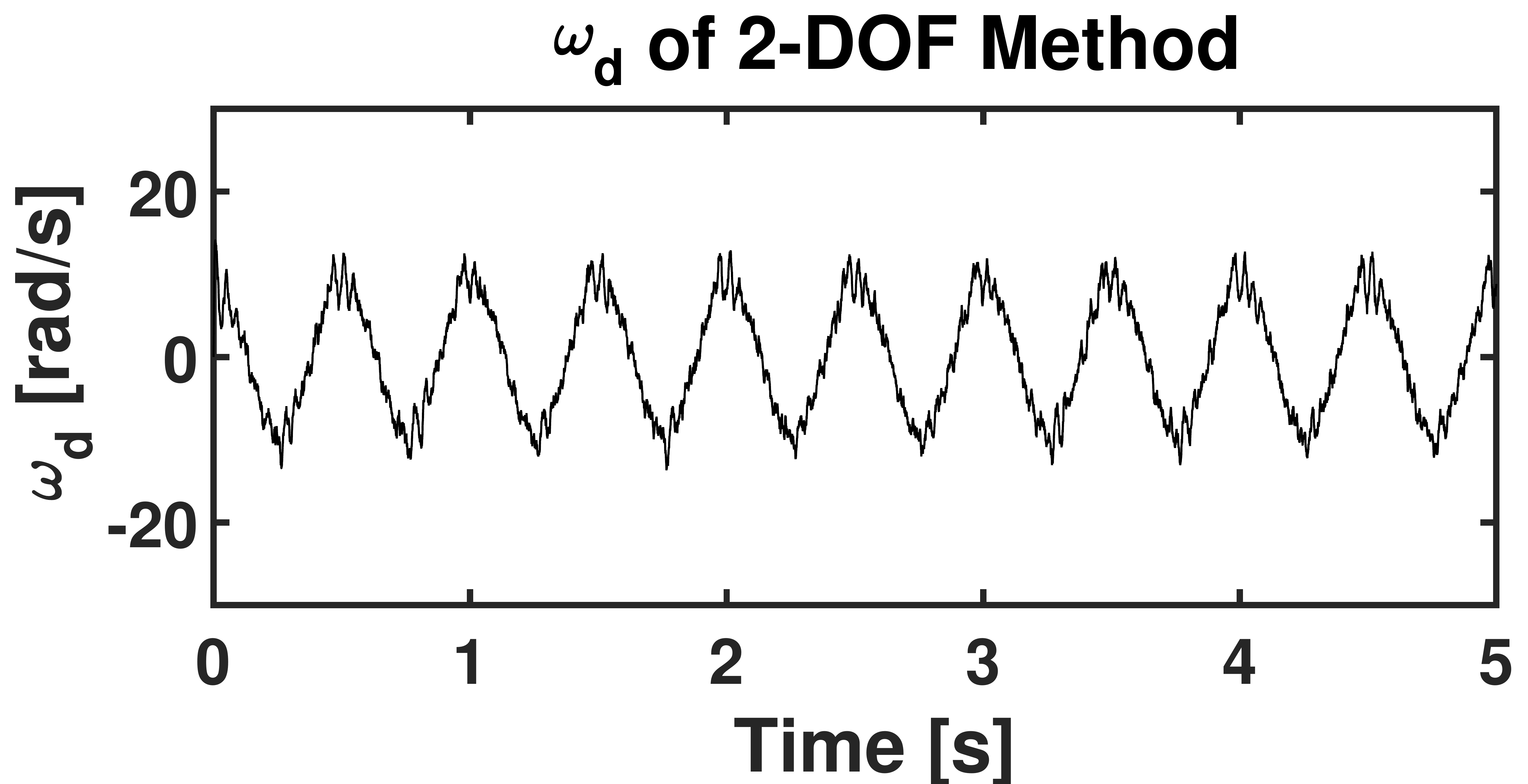}}
		\quad
		\subfigure[]{\includegraphics[width=0.47\columnwidth]{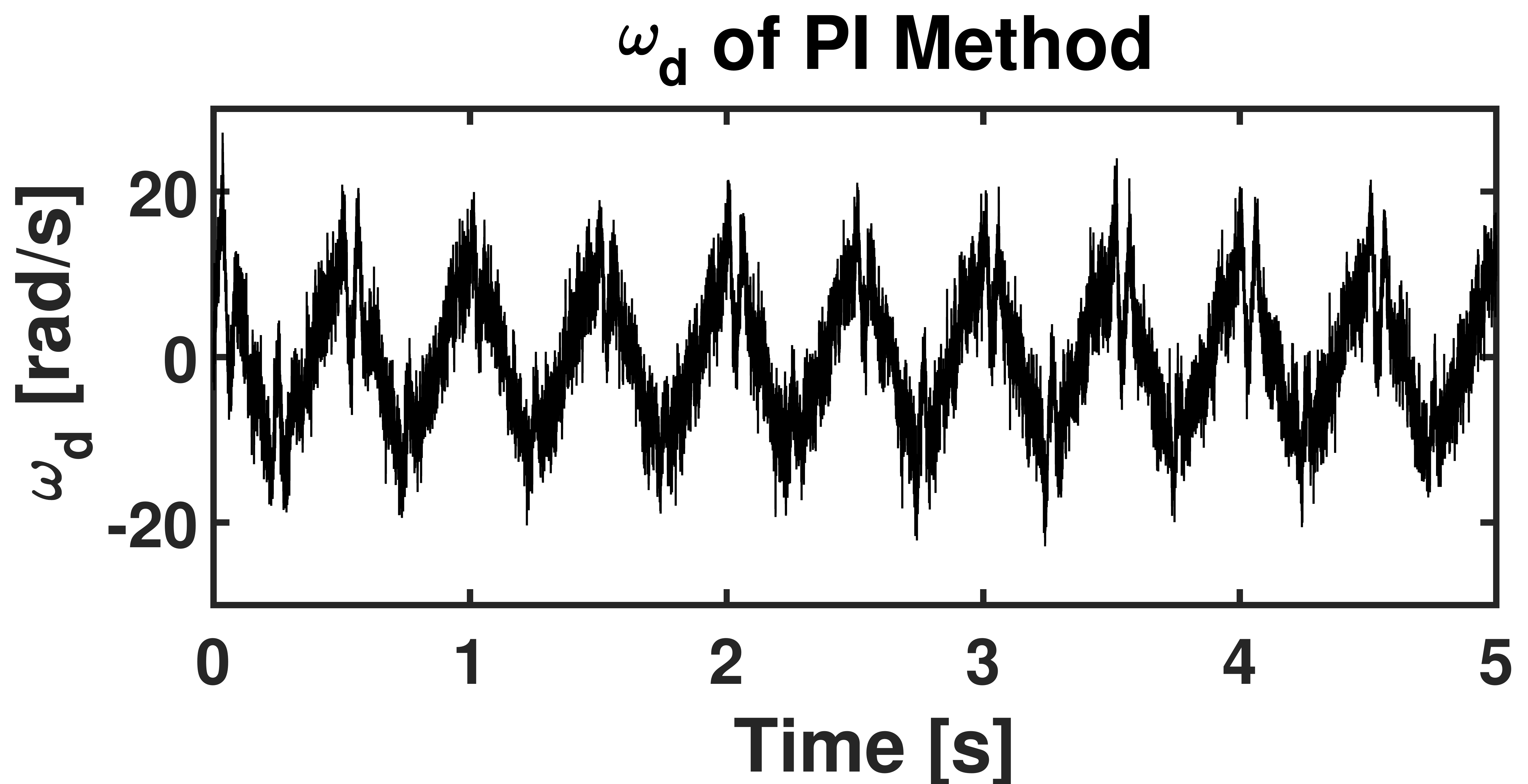}}
	\end{minipage}
	\caption{Experiment results of tracking a sinusoidal signal using the two methods. (b) and (c) are the plots of $\omega_d$ of the two methods respectively.}\label{exp_comparison_loadfixed}
\end{figure}

In the first experiment, the load was fixed at the system reference position. To compare the 2-DOF controller with a conventional one, a PI torque controller was tuned as $C_\tau(s)=204+111\frac{1}{s}$. The torque reference signal was set to a sinusoidal signal with a magnitude of 0.033 Nm and a frequency of 2 Hz, and an extra white Gaussian noise with a variance of 0.01 was added into the torque feedback channel. As shown in Fig.~\ref{exp_comparison_loadfixed}, both of the two control methods tracked the reference very well. However, the 2-DOF
controller reacted to noise more quickly and showed better noise rejection.
\begin{figure}[!htbp]
\centering
\includegraphics[width=\columnwidth]{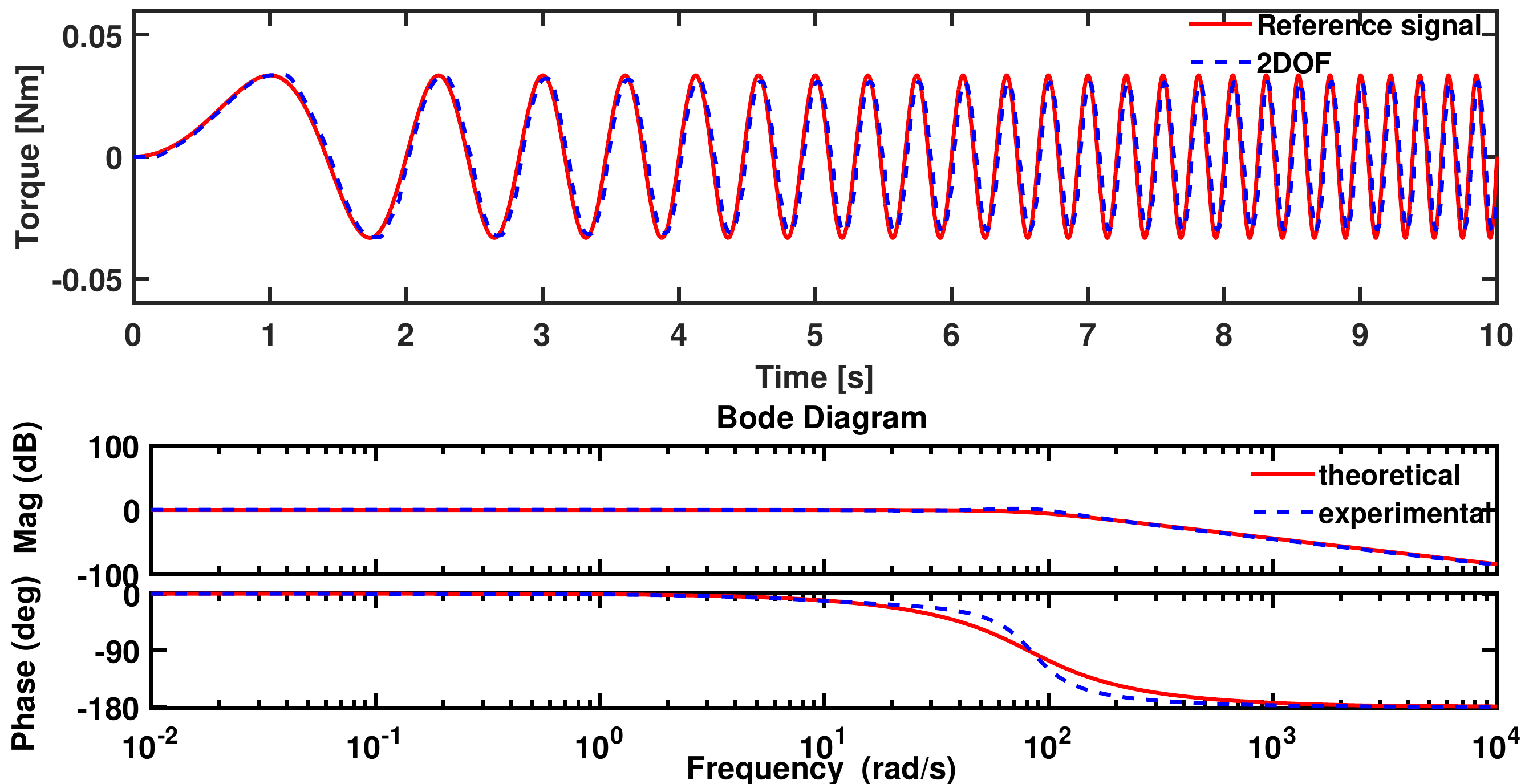}
\caption{Experimental results of tracking a chirp signal using the 2-DOF method. The bode plots of the estimated closed-loop transfer function and the theoretically expected frequency response are also illustrated.}\label{experiment_chirpsignal_bode_loadfixed}
\end{figure}
\begin{figure}[!htbp]
\centering
\includegraphics[width=\columnwidth]{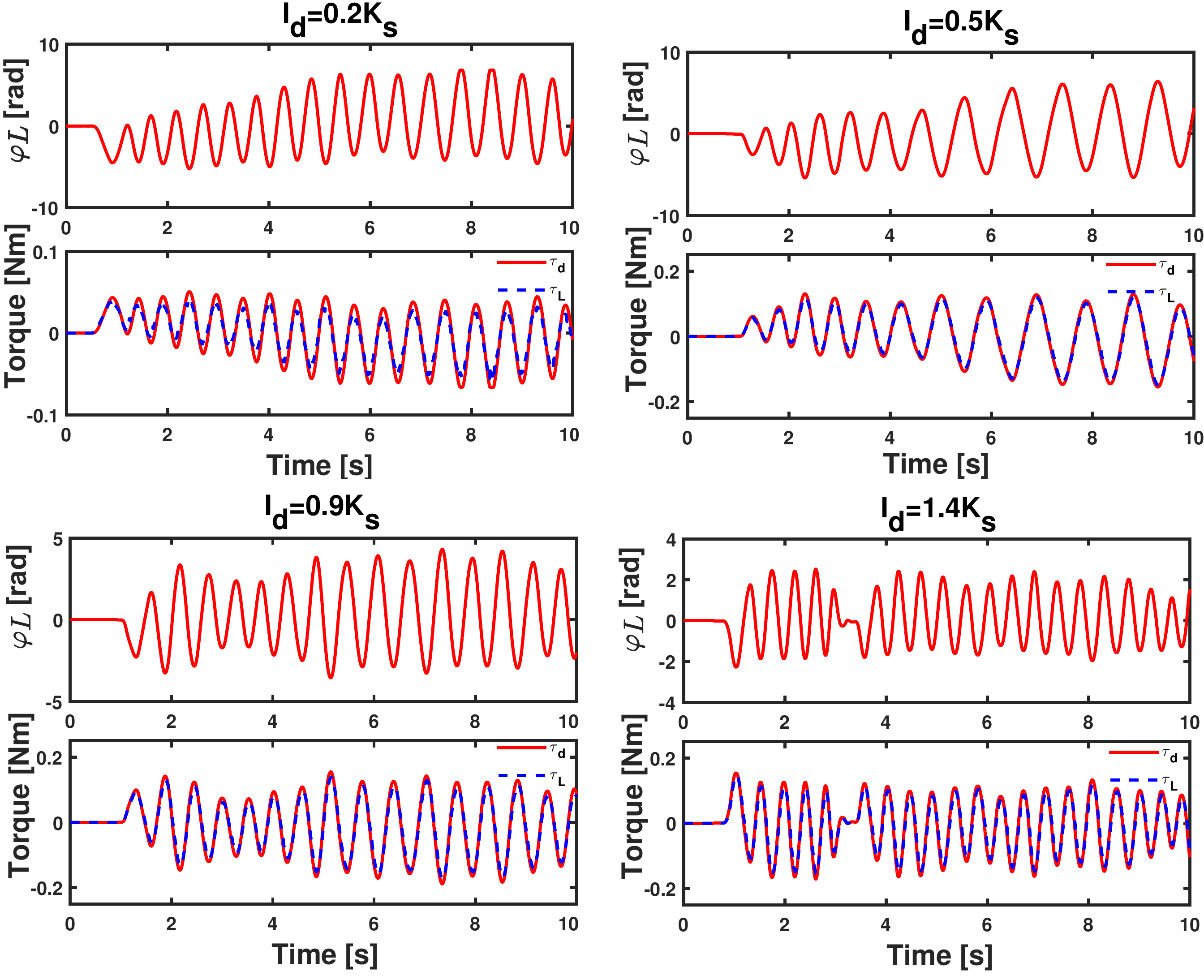}
\caption{Experimental results of impedance control in different virtual impedance.}\label{experiment_impedance}
\end{figure}
\begin{figure}[!htbp]
\centering
\includegraphics[width=0.9\columnwidth]{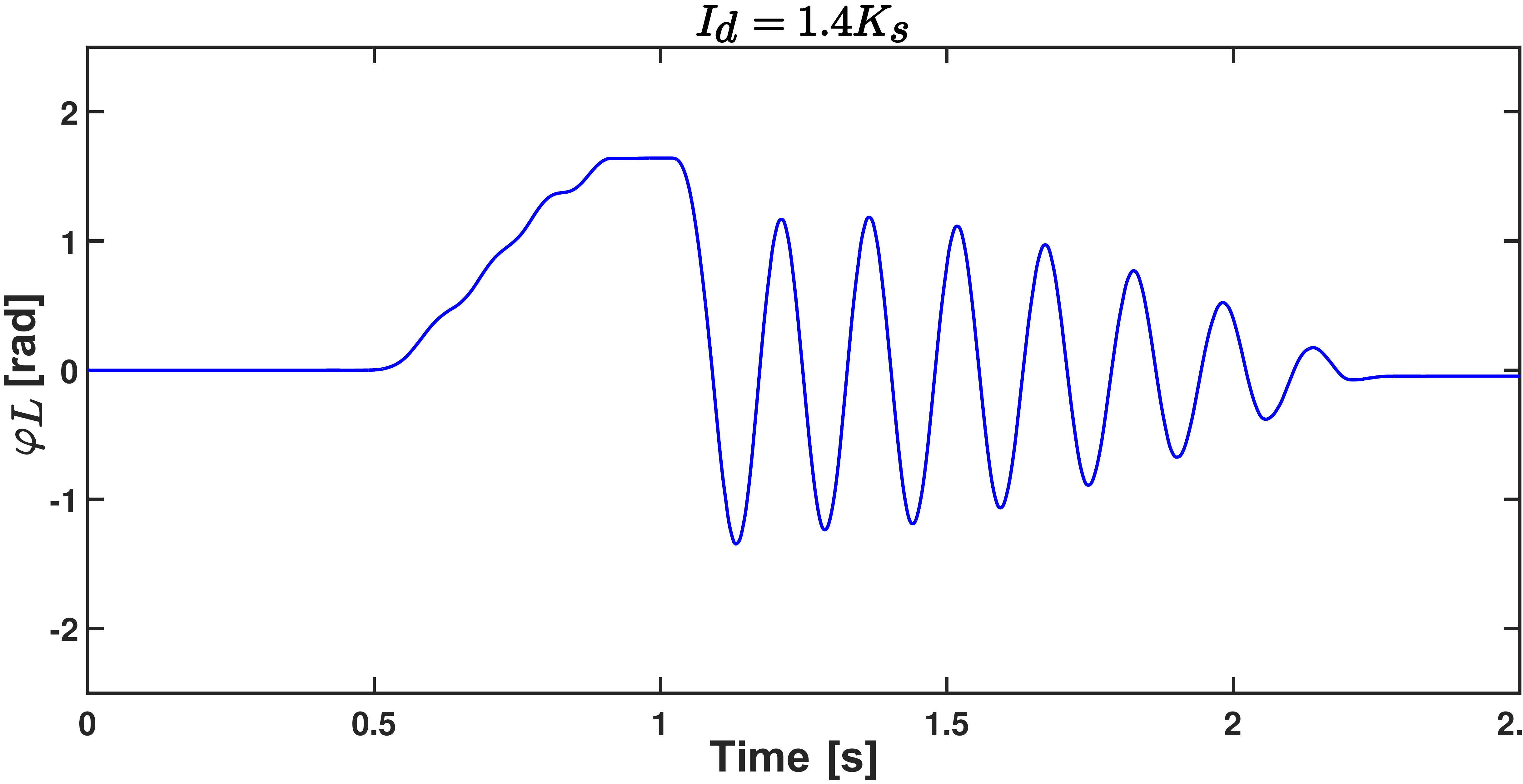}
\caption{An instantaneous load motion resulting in decaying oscillations.}\label{experiment_oscillation}
\end{figure}

In the second experiment, the torque reference signal was set to a chirp signal with frequencies ranging from 0 to 5Hz and a magnitude of 0.033 Nm. The tracking results of the 2-DOF method was illustrated in Fig.~\ref{experiment_chirpsignal_bode_loadfixed}. The bode plots of theoretical closed-loop transfer function and experimental estimated were plotted in Fig.~\ref{experiment_chirpsignal_bode_loadfixed}. A bandwidth of 17 Hz with a phase lag of ${130^ \circ }$ at this cutoff frequency was achieved.

In the third experiment, the handle was moved along the guide by human in different frequencies. The virtual stiffness changed from $0.2K_s$ to $1.4K_s$. The results were shown in Fig.~\ref{experiment_impedance}. 
Applying an instantaneous load motion to test the stability of the system, it resulted in decaying oscillations of the slider as shown in Fig.~\ref{experiment_oscillation}. It can be seen that with the designed 2-DOF force controller, the impedance controller and the compensator, virtual stiffness ranging from 0.2 to 1.4 times of the physical spring stiffness $K_s$ can be realized.

\section{Conclusion}\label{section5}

This paper has demonstrated the efficacy of the 2-DOF control approach for the challenging torque control and impedance control problem of a cable-driven SEA. The 2-DOF torque controller performed well in the presence of noise, disturbance, motor saturation and modeling uncertainties. Experimental results demonstrated that the 2-DOF controller brought better robustness than the PI controller. It is convenient to adjust these competing performances. Further, the impedance controller for human-robot interaction was designed and implemented with a torque compensator. Both
simulation and experiments have validated the efficacy of the 2-DOF control structure for the cable-driven SEA system.

\section*{Acknowledgments}

The authors would like to thank Prof. Dr. Li Qiu and Ms. Dan Wang from the Hong Kong University of Science and Technology for the helpful discussion, and Mr. Bin Li from Nankai University for the mechanical design of the prototype of the cable-driven SEA.

\bibliography{reference}

\begin{thebibliography}{10}
\providecommand{\url}[1]{#1}
\csname url@samestyle\endcsname
\providecommand{\newblock}{\relax}
\providecommand{\bibinfo}[2]{#2}
\providecommand{\BIBentrySTDinterwordspacing}{\spaceskip=0pt\relax}
\providecommand{\BIBentryALTinterwordstretchfactor}{4}
\providecommand{\BIBentryALTinterwordspacing}{\spaceskip=\fontdimen2\font plus
\BIBentryALTinterwordstretchfactor\fontdimen3\font minus
  \fontdimen4\font\relax}
\providecommand{\BIBforeignlanguage}[2]{{%
\expandafter\ifx\csname l@#1\endcsname\relax
\typeout{** WARNING: IEEEtran.bst: No hyphenation pattern has been}%
\typeout{** loaded for the language `#1'. Using the pattern for}%
\typeout{** the default language instead.}%
\else
\language=\csname l@#1\endcsname
\fi
#2}}
\providecommand{\BIBdecl}{\relax}
\BIBdecl

\bibitem{Pratt1995}
G.~A. Pratt and M.~M. Williamson, ``Series elastic actuators,'' in
  \emph{Proceedings of the IEEE/RSJ International Conference on Intelligent
  Robots and Systems}, vol.~1, 1995, pp. 399--406.

\bibitem{Robinson1999}
D.~W. Robinson, J.~E. Pratt, D.~J. Paluska, and G.~Pratt, ``Series elastic
  actuator development for a biomimetic walking robot,'' in \emph{Proceedings
  of the IEEE/ASME International Conference on Advanced Intelligent
  Mechatronics}, 1999, pp. 561--568.

\bibitem{Kong2012}
K.~Kong, J.~Bae, and M.~Tomizuka, ``A compact rotary series elastic actuator
  for human assistive systems,'' \emph{IEEE/ASME Transactions on Mechatronics},
  vol.~17, no.~2, pp. 288--297, 2012.

\bibitem{Yu2013}
H.~Yu, S.~Huang, N.~V. Thakor, G.~Chen, S.~L. Toh, M.~S. Cruz, Y.~Ghorbel, and
  C.~Zhu, ``A novel compact compliant actuator design for rehabilitation
  robots,'' in \emph{Proceedings of the IEEE International Conference on
  Rehabilitation Robotics (ICORR)}, 2013, pp. 1--6.

\bibitem{Chapuis2006}
D.~Chapuis, R.~Gassert, G.~Ganesh, E.~Burdet, and H.~Bleuler, ``Investigation
  of a cable transmission for the actuation of \textsc{MR} compatible haptic
  interfaces,'' in \emph{Proceedings of the First IEEE/RAS-EMBS International
  Conference on Biomedical Robotics and Biomechatronics}, 2006, pp. 426--431.

\bibitem{Caverly2015}
R.~J. Caverly, J.~R. Forbes, and D.~Mohammadshahi, ``Dynamic modeling and
  passivity-based control of a single degree of freedom cable-actuated
  system,'' \emph{IEEE Transactions on Control Systems Technology}, vol.~23,
  no.~3, pp. 898--909, 2015.

\bibitem{Veneman2006}
J.~F. Veneman, R.~Ekkelenkamp, R.~Kruidhof, F.~C. V.~D. Helm, and H.~V.~D.
  Kooij, ``A series elastic-and bowden-cable-based actuation system for use as
  torque actuator in exoskeleton-type robots,'' \emph{The International Journal
  of Robotics Research}, vol.~25, no.~3, pp. 261--281, 2006.

\bibitem{Oblak2011}
J.~Oblak and Z.~Matja\u{c}i\'{c}, ``Design of a series visco-elastic actuator
  for multi-purpose rehabilitation haptic device,'' \emph{Journal of
  neuroengineering and rehabilitation}, vol.~8, no.~1, p.~3, 2011.

\bibitem{Sergi2015}
F.~Sergi and M.~K. O'Malley, ``On the stability and accuracy of high stiffness
  rendering in non-backdrivable actuators through series elasticity,''
  \emph{Mechatronics}, vol.~26, pp. 64--75, 2015.

\bibitem{Robinson2000}
D.~W. Robinson, ``Design and analysis of series elasticity in closed-loop
  actuator force control,'' Thesis, Massachusetts Institute of Technology,
  2000.

\bibitem{Vallery2008}
H.~Vallery, J.~Veneman, E.~V. Asseldonk, R.~Ekkelenkamp, M.~Buss, and H.~V.~D.
  Kooij, ``Compliant actuation of rehabilitation robots,'' \emph{Robotics and
  Automation Magazine}, vol.~15, no.~3, pp. 60--69, 2008.

\bibitem{Wyeth2008}
G.~Wyeth, ``Demonstrating the safety and performance of a velocity sourced
  series elastic actuator,'' in \emph{Proceedings of the IEEE International
  Conference on Robotics and Automation (ICRA)}, 2008, pp. 3642--3647.

\bibitem{Yoo2015}
S.~Yoo and W.~K. Chung, ``\textsc{SEA} force/torque servo control with
  model-based robust motion control and link-side motion feedback,'' in
  \emph{Proceedings of the IEEE International Conference on Robotics and
  Automation (ICRA)}, 2015, pp. 1042--1048.

\bibitem{Lu2015}
J.~Lu, K.~Haninger, W.~Chen, and M.~Tomizuka, ``Design and torque-mode control
  of a cable-driven rotary series elastic actuator for subject-robot
  interaction,'' in \emph{Proceedings of the IEEE International Conference on
  Advanced Intelligent Mechatronics (AIM)}, 2015, pp. 158--164.

\bibitem{Horowitz1963}
I.~M. Horowitz, \emph{Synthesis of feedback systems}.\hskip 1em plus 0.5em
  minus 0.4em\relax London: Acdemic Press, 1963.

\bibitem{Youla1985}
D.~C. Youla and J.~J. Bongiorno~Jr, ``A feedback theory of
  two-degree-of-freedom optimal wiener-hopf design,'' \emph{IEEE Transactions
  on Automatic Control}, vol.~30, no.~7, pp. 652--665, 1985.

\bibitem{Qiu2010}
L.~Qiu and K.~Zhou, \emph{Introduction to feedback control}.\hskip 1em plus
  0.5em minus 0.4em\relax New Jersey: Pearson Prentice Hall, 2010.

\bibitem{Huang2015}
P.~Huang, ``Theoretical and experimental studies on two-degree-of-freedom
  controllers,'' Thesis, Hong Kong University of Science and Technology, 2015.

\bibitem{ZouWulin2016}
W.~Zou, W.~Tan, Z.~Yang, and N.~Yu, ``Torque control of a cable-driven series
  elastic actuator using the 2-\textsc{DOF} method,'' in \emph{Proceedings of
  the 35th Chinese Control Conference}, 2016, pp. 6239--6243.

\bibitem{Yu2014}
N.~Yu, W.~Tan, and J.~Liu, ``Design and analysis of a wrist-hand manipulator
  for rehabilitation of upper limb dexterous function,'' in \emph{Proceedings
  of the IEEE International Conference on Robotics and Biomimetics (ROBIO)},
  2014, pp. 797--801.

\end{thebibliography}

\end{document}